\documentclass[twocolumn,aps,showpacs]{revtex4}
\input epsf
\begin{document}
%\draft
\title{Power law relaxation in a complex system: 
Omori law after a financial market crash}
\author{F. Lillo$^{*}$ and R. N. Mantegna$^{*,\dag}$\ }
\affiliation{$^{*}$ Istituto Nazionale per la Fisica della Materia, Unit\`a di Palermo, 
Viale delle Scienze, I-90128, Palermo, Italia\\
 $^{\dag}$  Dipartimento di Fisica e Tecnologie Relative, 
Universit\`a di Palermo, Viale delle Scienze, I-90128
Palermo, Italia.}

\begin{abstract}
We study the relaxation dynamics of a financial market
just after the occurrence of a crash
by investigating the number of times
the absolute value of 
an index return 
is exceeding a given threshold value. 
We show that the empirical observation of a power law evolution 
of the number of events exceeding the selected threshold 
(a behavior known as the Omori law in geophysics)
is consistent with the simultaneous occurrence 
of (i) a return probability density function characterized 
by a power law asymptotic behavior and
(ii) a power law relaxation decay of its typical scale.
Our empirical observation cannot be explained within
the framework of simple and widespread stochastic volatility models.
\end{abstract}

\maketitle

Several complex systems are statistically characterized 
by power-law distributions. Examples are earthquakes, 
financial markets, landslides, forest fires 
and scale free networks. Power law distributions
imply that
rare events 
are occurring with a finite non-negligible probability in complex systems. 
It is therefore meaningful to ask the following scientific question: 
how is the dynamics of a complex system affected when the system 
undergoes to an extreme event? An answer to this question concerning
earthquakes was provided by Omori
more than a century ago \cite{Omori1894}. 
The Omori law describes the non stationary period observed after a 
big earthquake. In his study, 
the number of aftershocks per unit of time is 
described by a power law and a time scale
for the relaxation process of the complex system to its typical state 
does not exist. Non exponential relaxation to a typical state 
has also been observed in several physical and social systems. 
For example, power law relaxation has been theoretically
predicted and experimentally observed in spin glasses \cite{Bouchaud1992},
condensed matter systems \cite{chamberlin}, 
microfracturing phenomena \cite{zapperi},
physical systems described by a fractional Fokker-Planck 
equation \cite{Metzler1999}, 
in the kinetics of reversible bimolecular reactions 
\cite{Gopich00}, in two-dimensional arrays of 
magnetic dots interacting by long-range dipole-dipole interactions
\cite{Sampaio01},  in the Internet dynamical response
\cite{johansen2001} and in the Internet traffic \cite{abe}. 
 
In the present study we investigate the dynamics of a model
complex system when it is moved far away from its typical state by 
the occurrence of an extreme event.
This is done by investigating the statistical properties of time 
series of financial indices in the time period immediately
after a financial crash. These market phases are indeed  
strongly non-stationary and we show that a time power law 
relaxation is detected when the financial market is moved far 
away from its typical behavior.

Financial time series of stock or index returns are modeled  
in terms of random processes \cite{Samuelson65,Cootner64}.
Empirical investigations show that the time series of stock or
index return is not strictly sense stationary. In fact the 
volatility of the financial asset,
i.e. the standard deviation of asset returns describing
the typical scale of the process, 
is itself a stochastic process fluctuating in time 
\cite{Hull,Campbell97}.  
The non-stationary evolution of asset returns can sometimes show 
relaxation time patterns. 
Specifically, decaying patterns of volatility  
are observed in time periods immediately after a 
financial crash. An illustrative example of such non-stationary
time pattern is given in Fig. 1 where we plot
the one-minute logarithm changes of the index $r(t)$
(a quantity essentially equivalent to return)
for the Standard and Poor's 500 (S\&P500)
index during 100 trading days after the Black
Monday (19 October 1987). 
The pattern observed in Fig. 1 is not invariant under
time-reversal.
Other examples of statistical properties of market 
which are not time-reversal have been observed in 
the investigation of cross-sectional quantities 
computed for a set of stocks before and after 
financial crashes \cite{Lillo2001}. 

A direct characterization of the time evolution of the scale 
of the random process of return is extremely difficult in 
financial markets and in several other complex systems due to
the fact that the random variable is highly fluctuating and 
that system is unavoidably monitored by just 
recording a single random realization. We make use of a different
and statistically more robust method. 
Specifically,
we quantitatively characterize the time series of index returns
in the non-stationary time period by investigating the number
of times $|r(t)|$ is exceeding a given threshold value. This 
investigation is analogous to the investigation of the number $n(t)$
of aftershock earthquakes measured at time $t$ after the main 
earthquake. The Omori law $n(t) \propto t^{-p}$ says 
that the number of aftershock earthquakes per unit time measured at
time $t$ after the main earthquake decays as a power law.
In order to avoid divergence at $t=0$ Omori law is often
rewritten as
\begin{equation}
n(t)=K(t+\tau)^{-p}
\end{equation}
where $K$ and $\tau$ are two positive constants.
An equivalent formulation of the Omori law more suitable for
comparison with real data can be obtained by integrating
equation (1) between $0$ and $t$. In this way the cumulative 
number of aftershocks observed until time $t$ after the main 
earthquake is
\begin{equation}
N(t)=K[(t+\tau)^{1-p}-\tau^{1-p}]/(1-p)
\end{equation}
when $p \ne 1$ and $N(t)=K\ln(t/\tau+1)$
for $p=1$. The value of the exponent $p$ for earthquakes ranges
between $0.9$ and $1.5$. Because $N(t)$ is related
to $n(t)$ by a summation, the fluctuation in $N(t)$ is substantially
reduced compared with the fluctuation in $n(t)$.  Hence customary 
measurement of $N(t)$ leads to a more reliable characterization
of the aftershock period than measurement of $n(t)$.

We first investigate the index returns during the time period 
after the Black Monday crash (19 October 1987)
occurred at New York Stock Exchange 
(NYSE). This crash was one of the worst crashes occurred in the entire 
history of NYSE. The S\&P500 
went down $20.4\%$
that day. In our investigation, we select a 60 day after crash
time period ranging from 20 October 1987 to 14 January 1988.
This time period is chosen to maximize the time period investigated
by simultaneously ensuring that the relaxation process is still 
going on.
The selected value is not a critical one and time windows of 50 or 70
trading days provide similar results.
For the selected time period, we investigate the one-minute return time
series of the S\&P500 Index. The first estimate concerns the 
unconditional one-minute volatility which is equal to 
$\sigma=4.91 \times 10^{-4}$.
In Fig. 2 we show the cumulative number of events $N(t)$
detected by considering all the occurrences observed when the the 
absolute value of index return exceeds a threshold value $\ell$
chosen as $4 \sigma$, $5 \sigma$, $6 \sigma$ and $7 \sigma$.
For all the selected threshold values we observe a 
nonlinear behavior. Nonlinear
fits performed with the functional form of equation (2) 
well describes the empirical data for the entire time period.  
This paradigmatic behavior is 
not specific of the Black Monday 
crash of the S\&P 500 index. In fact, we observe 
similar results also for a 
stock price index weighted by market 
capitalization for the time periods 
occurring after the 27 October 1997 and the 
31 August 1998 stock market crashes. 
This index has been computed  
selecting the 30 most capitalized stocks 
traded in the NYSE and by using the high-frequency
data of the {\it Trade and Quote} database 
issued by the NYSE.
In Fig. 3 we show $N(t)$ for $\ell=4\sigma$
where $\sigma$ is again the unconditional one-minute volatility 
in the considered periods. We estimate $\sigma=4.54 \times 10^{-4}$ 
during the period from 28 October 1997
to 23 January 1998 and $\sigma=6.09 \times 10^{-4}$ 
during the period from 1 September 1998 to 24 November 1998.
In the left part of Table I, we summarize the values of the $p$ 
exponents obtained by best fitting with equation (2) the 
cumulative number of events exceeding the selected threshold 
values for the considered market crashes. 
The value of the exponent $p$ varies in the interval between 0.70 
and 0.99 . The estimate of the exponent $p$ is slightly 
increasing when the threshold value $\ell$ is increasing.  
Below we will comment the relation between this observation
and the properties of the index return 
probability density function (pdf).
The detected nonlinear behavior of $N(t)$ is specific to
aftercrash market period. In fact an approximately linear behavior of 
$N(t)$ is observed when a market period of roughly
constant volatility such as, for example, the 1984 year 
is investigated. This is due to the fact that 
when the process is stationary the frequency of 
aftershock $n(t)$ is on average constant in time and 
therefore the 
cumulative number $N(t)$ increases linearly in time.
In terms of equation (2)
this implies that the exponent $p$ is equal to zero.
For independent identically distributed random time series
it is possible to characterize $n(t)$ in terms of an 
homogeneous Poisson process \cite{Embrechts}. The results summarized in 
the left part of Table I imply 
that the time period immediately after a big market crash 
has statistical properties which are different from 
constant volatility periods. In particular, 
index return cannot be modeled in terms of independent 
identically distributed random process after a big market crash.

\begin{table*}
\caption{Exponents obtained from the empirical analyses of 60 day
market periods occurring after the 19 October 1987, 27 October 1997 and
31 August 1998 market crashes.} 
\begin{ruledtabular} 
\begin{tabular}{c|cccc|ccc}
 &\multicolumn{4}{c|}{$p$}&\multicolumn{1}{c}{$\alpha$}&\multicolumn{1}{c}{$\beta$}&
 \multicolumn{1}{c}{$\alpha~\beta$}\\
 ~& $4~\sigma$ & $5~\sigma$ & $6~\sigma$ & $7~\sigma$ & ~ & ~ & 
 ~ \\ \hline
 1987 & 0.85 &  0.90 &  0.99 &  0.99 & $3.18 \pm 0.34 $ & 
 $0.32 \pm 0.02$ &$1.02\pm 0.13$ \\
  1997 & 0.70 &  0.73 &  0.73 &  0.76& $3.67 \pm 0.40 $ & 
 $0.22 \pm 0.04$ & $0.81\pm 0.17$ \\
  1998 & 0.99 & 0.99  & 0.99 & 0.99 & $3.49 \pm 0.37 $ & 
 $0.32 \pm 0.05$ & $1.12\pm 0.21$ \\
\end{tabular}
\end{ruledtabular} 
\end{table*}

%\section{Model of the Aftercrash Period}

%\subsection{The Model} 

The empirical evidence of the power-law decrease of the 
frequency of aftershocks is consistent with a power-law 
decay of volatility after a major crash. In order to prove
this claim, we describe
the empirical behavior of $N(t)$ by assuming that during 
the time period after a big crash the stochastic variable $r(t)$ 
is the product of a time dependent 
scale $\gamma(t)$ times a stationary stochastic process $r_s(t)$.
For the sake of simplicity, we also assume that the pdf
of $r(t)$ is approximately symmetrical.
Under these assumptions, the frequency of events 
of $|r(t)|$ larger than $\ell$ observed at time $t$ is 
\begin{equation} 
n(t) \propto 2\int_{\ell}^{+\infty} f(r,t)dr
\end{equation}
where $f(r,t)$ is the pdf of $r(t)$ at time $t$.
One can rewrite equation (3) in terms of the cumulative
distribution function $F_s(r_s)$ of the random variable 
$r_s(t)$ as
\begin{equation} 
n(t) \propto 1-F_s\left(\ell/\gamma(t)\right).
\end{equation}

In this description, the specific form of the time evolution of 
$n(t)$ is, for large values of the 
threshold $\ell$, controlled by the properties of
(i) the time evolution of the scale $\gamma(t)$ and 
(ii) the asymptotic behavior of the pdf for large 
values of $|r_s(t)|$. 

By assuming that the stationary return pdf behaves 
asymptotically as a power law 
\begin{equation} 
f_s(r_s)\sim \frac{1}{r_s^{\alpha+1}},
\end{equation}
the frequency of events $n(t)$ becomes for large values of $\ell$
\begin{equation} 
n(t) \sim \left(\gamma(t)/\ell\right)^{\alpha}.
\end{equation}

By hypothesizing that $\gamma(t)\sim \exp{(-kt)}$,
the frequency of events above threshold is expected
exponentially decreasing $n(t)\sim \exp{(-\alpha kt)}$. 
Conversely, when the scale of the stochastic process
decays as a power law
$\gamma(t)\sim t^{-\beta}$, the frequency of events above threshold is
power law decaying as $n(t)\sim 1/t^p$. It is worth noting
that the exponent $p$ is given by 
\begin{equation} 
p=\alpha~\beta.
\end{equation}
The previous relation links the exponent $p$ governing the
number of events exceeding a given threshold to the
$\alpha$ exponent of the power law return cumulative distribution
and to the $\beta$ exponent of the power law decaying
scale. 
It is worth noting that a power law behavior
of the return pdf is observed only for
large absolute values of returns. Hence, the relation between
exponents (eq. (7)) is valid 
only for large values of the threshold $\ell$ used to determine 
the exponent $p$. Our theoretical considerations 
show that a number of events
above threshold decaying as a power law, i.e. the analogous
of the Omori law, is consistent with the simultaneous occurrence 
of: (i) a return pdf characterized 
by a power law asymptotic behavior and
(ii) a non-stationary time evolution of the return pdf 
whose scale is decaying in time as a power law. These hypotheses
are consistent with recent empirical results. In fact, a return pdf
characterized by a power law asymptotic behavior has been
observed in the price dynamics of several stocks \cite{Lux96,Gopi98}.
To the best of our knowledge the only investigation on the 
decay of volatility after a crash has been performed in 
Ref. \cite{Sornette96} where
a power law or power law log-periodic decay of implied volatility has 
been observed in the S\&P500 after the 1987 financial
crash. We would like to stress that implied volatilty is different
from our $\gamma(t)$ because implied volatiltiy is obtained 
from index derivative prices by using the Black and Scholes formula
instead that directly from data.
Moreover the value of the exponent governing the decay of volatility
is different in our study and in Ref. \cite{Sornette96}.
The analytical considerations developed above indicate that stochastic
volatility models of price dynamics are able to describe the behavior
of an index after a crash when they predict the volatility power-law 
decay in time after a crash. Therefore simple autoregressive models, 
such as GARCH(1,1) \cite{Bollerslev86} models, are unable to 
describe the observed behavior.
GARCH processes in their most compact
form cannot show a scale of the stochastic process decaying
as a power law after a big event. 
By analytical calculation and by performing numerical simulations 
we have shown that these models are characterized by
an exponential decay of the scale of the process \cite{jedc}.

In order to show that empirical data are consistent with our description
of aftershock periods, we empirically study the 
time evolution of the scale of the process. To this
end, by using the ordinary least square method,
we fit the absolute value 
of return with the functional form $f(t)=c_1 t^{-\beta} +c_2$
in the 60 days after each considered 
market crash. We check that the relation
$c_1 t^{-\beta} >>c_2$ is verified in the
investigated period. 
The best estimation of $c_1$ and $c_2$ are $6.3~10^{-4}$ and
$2.8~10^{-6}$ for the 1987 crash, $5.1~10^{-4}$ and $4.3~10^{-5}$
for the 1997 crash and $4.4~10^{-4}$ and $1.0~10^{-4}$ for the
1998 crash. The time $t$ is expressed in trading day. 
By using the relation, $r(t)=\gamma(t) r_s(t)$,
the $\beta$ exponent obtained
is also the exponent controlling the scale $\gamma(t)$.
In order to estimate the $\alpha$ exponent governing
the stationary part of the return evolution we define 
a new variable $r_p(t)$ obtained dividing $r(t)$ by the moving
average of its absolute value. The averaging window is
set to 500 trading minutes. The quantity $r_p(t)$ is a proxy for the
stationary return $r_s(t)$. 
We investigate the asymptotic properties for large absolute 
values of the stochastic process $r_p(t)$ by computing the 
Hill's estimator \cite{Hill} of the process computed over the largest 
1\% values of $|r_p(t)|$. To assess the
reliability of the $\alpha$ estimate obtained with this
method we also compute its 95\% confidence interval.
The 95\% confidence interval is obtained by computing 
$C_{95} \alpha /\sqrt{m}$ where $C_{95}$ is the value
at which the normal distribution is equal to 0.95 and
$m$ is the number of records located in the distribution
tail.
With our procedure, we obtain a value of the exponent 
$\alpha$ which is ranging from 3.18 to 3.67. These 
values are consistent
with the observations performed by different authors on
the power law behavior governing large absolute returns 
in stocks and stock indices \cite{Lux96,Gopi98}.

The estimates of $\alpha$ and $\beta$ values 
are shown in the right part of Table I for all the 
investigated market crashes. The last column of the Table
gives the value of the product $\alpha~\beta$ that is to
be compared with the values of $p$ summarized in the left 
part of the Table. The agreement is increasingly good 
for values of $p$ obtained for large values of the threshold. 
This is expected because only for large threshold the
relevant part of the return pdf is well described by a 
power law behavior.

Finally, we investigate the properties of $N(t)$ computed 
for the random variable $r_p (t)$. This variable is our 
proxy for $r_s (t)$
and therefore a linear behavior of $N(t)$ is expected
for each value of the threshold chosen. 
>From our definition of $r_p (t)$, it follows that the mean of the
absolute value of $r_p (t)$ is equal to one.
In Fig. 4, we show
$N(t)$ for the market crash of 19 October 1987 when
$\ell$ is ranging from $4$ to $13$.
For all values of the threshold, $N(t)$ is approximately linear showing that
$r_p(t)$ provides a good proxy for $r_s(t)$. 
Moreover, starting from equation (4), one can show that 
the slope $\eta$ of $N(t)$ is proportional to the quantity
$1-F_s(\ell)$. We determine
$\eta$  with a best linear fit of $N(t)$ for each value
of $\ell$. The results are shown in the inset of Fig. 4.
Under the assumption of equation (5), the expected relation
between $\eta$ and $\alpha$ is $\eta \sim \ell^{-\alpha}$.
The inset also shows our best fit of $\eta$ with a power law
relation as a solid line. The best fitting exponent is $\alpha=3.14$ 
when $\ell \ge 7$. This value of $\alpha$ is consistent 
with the value obtained with the Hill estimator (see Table I).

%\section{Discussion and Conclusions}

In conclusion our results show that time periods of the order of 60 
trading days (approximately 3 months in calendar time) occurring after a major 
financial crash can be modeled in terms of a new stylized statistical law.
Specifically, the number of index returns computed at a given time 
horizon occurring above a large threshold is well described by a power law
function which is analogous to the Omori law of geophysics.
 
The presence of a power law relaxation seems to be a common
behavior observed in a wide range of complex systems.
One possibility for this common occurrence is that the Omori law 
is a phenomenological manifestation of underlying 
common microscopic mechanisms governing the dynamics of complex systems
after an extreme event. An example of such mechanisms has been
proposed to model the magnetization relaxation in spin glasses 
where it has been shown that the presence of many metastable states 
whose lifetime are distributed according to a broad, power law
distribution implies a power law decay
of the magnetization during aging \cite{Bouchaud1992}.

\acknowledgments
Authors wish to thank INFM, ASI, MIUR and MIUR-FIRB
research projects for financial support. F.L. thanks
Vittorio Loreto for introducing him to the Omori law.

\newpage

\begin{figure}[t]
\epsfxsize=7cm
\centerline{\epsfbox{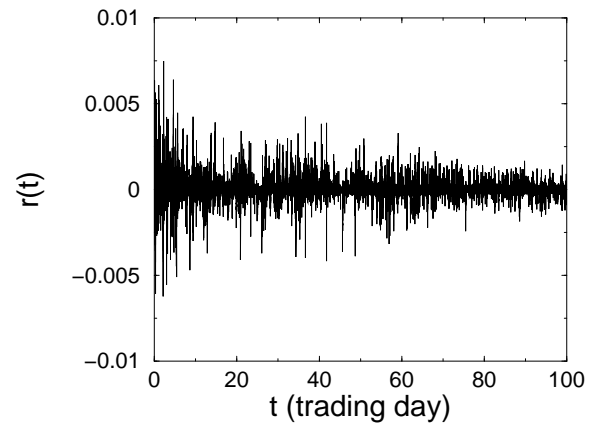}}
\caption{One-minute change of the natural logarithm 
of the Standard and Poor's
500 index during the 100 trading day time period immediately after the
Black Monday financial crash (20 October 1987 - 11 March 1998). 
A decrease of the 
typical scale of the stochastic process (volatility in the financial 
literature) is manifest making the stochastic process non-stationary.}
\label{fig1}
\end{figure}

%\newpage

\begin{figure}[t]
\epsfxsize=7cm
\centerline{\epsfbox{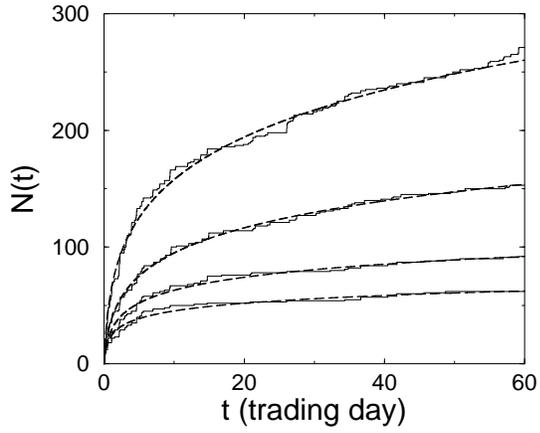}}
\caption{Cumulative number $N(t)$ of the number of times $|r(t)|$ is
exceeding a threshold $\ell$ during the 60 
trading days immediately after the Black Monday financial crash.
>From top to bottom we show the curves for values of $\ell$
equal to $4\sigma$, $5\sigma$, $6\sigma$ and $7\sigma$,
respectively.
The parameter $\sigma$ is the standard
deviation of the process $r(t)$ computed over the entire investigated period.
The dashed lines are best fits of equation (2).}
\label{fig2}
\end{figure}

%\newpage

\begin{figure}[t]
\epsfxsize=7cm
\centerline{\epsfbox{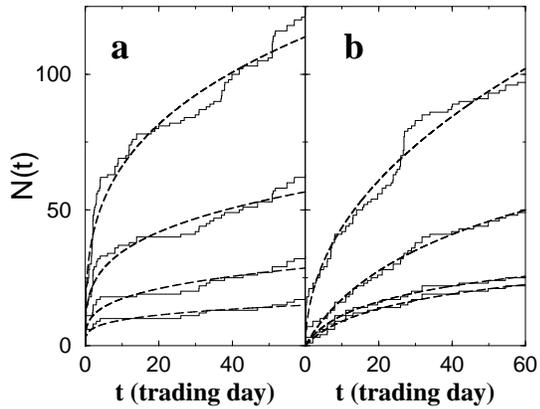}}
\caption{Cumulative number $N(t)$ of the number of times $|r(t)|$ is
exceeding the threshold $\ell$ during the 60 
trading days immediately after (a) the 27 October 1997 
and (b) the 31 August 1998 financial crashes.
In both panels, from top to bottom we show the curves 
for values of $\ell$
equal to $4\sigma$, $5\sigma$, $6\sigma$ and $7\sigma$,
respectively.
The parameter $\sigma$ is the standard
deviation of the process $r(t)$ computed over the entire investigated period.
The dashed lines are best fits of equation (2).}
\label{fig3}
\end{figure}

%\newpage

\begin{figure}[t]
\epsfxsize=7cm
\centerline{\epsfbox{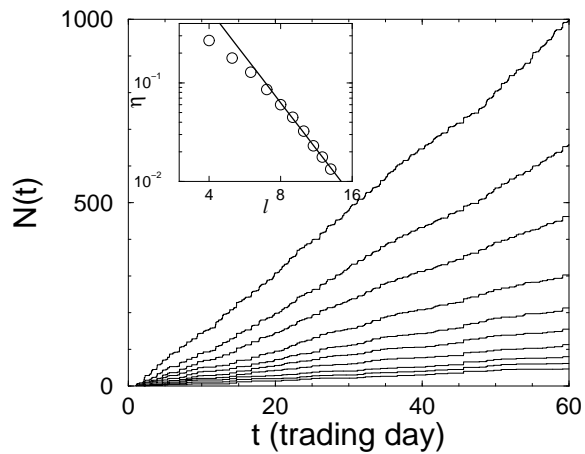}}
\caption{Cumulative number $N(t)$ of the number of times $|r_p(t)|$ is
exceeding a threshold $\ell$.
The data refer to the S\&P 500 index just after the 1987 crash.
$N(t)$ is computed for different values of the threshold $\ell$ 
ranging from 4 to 13. A linear behavior of $N(t)$ is observed for
all values of $\ell$. In the inset, we show the values of the
slope $\eta$ as a function of $\ell$ in a log-log plot. These 
values are computed by performing a best linear fitting of 
$N(t)$. The continuous line is the best fit of $\eta(\ell)$
with a power law behavior. The best fitting exponent is 3.14.}
\label{fig4}
\end{figure}

\end{document}